\documentclass[
reprint,
superscriptaddress,
 amsmath,amssymb,
aip,
floatfix,
aipsamp,
jmp
]{revtex4-1}
\usepackage{dcolumn}% Align table columns on decimal point
\usepackage{bm}% bold math
\usepackage{turnstile}

\usepackage{graphicx}
\usepackage{amsmath,amssymb}
\usepackage{url}
\usepackage{multirow}

\begin{document}

\title{Interfacial roughness and proximity effects in superconductor/ferromagnet CuNi/Nb heterostructures}
\author{Yu.~Khaydukov}
\affiliation{Max-Planck-Institut f\"ur Festk\"orperforschung, Heisenbergstra\ss e 1, D-70569 Stuttgart, Germany}
\affiliation{Max Planck Society Outstation at the Heinz Maier-Leibnitz Zentrum (MLZ), D-85748 Garching, Germany}
\affiliation{Skobeltsyn Institute of Nuclear Physics, 119991, Moscow State University, Moscow, Russia}
\author{R.~Morari}
\affiliation{Institute of Electronic Engineering and Nanotechnologies ASM, MD2028, Kishinev, Moldova}
\affiliation{Solid State Physics Department, Kazan Federal University, 420008 Kazan, Russia}
\author{O.~Soltwedel}
\affiliation{Max-Planck-Institut f\"ur Festk\"orperforschung, Heisenbergstra\ss e 1, D-70569 Stuttgart, Germany}
\affiliation{Max Planck Society Outstation at the Heinz Maier-Leibnitz Zentrum (MLZ), D-85748 Garching, Germany}
\author{T.~Keller}
\affiliation{Max-Planck-Institut f\"ur Festk\"orperforschung, Heisenbergstra\ss e 1, D-70569 Stuttgart, Germany}
\affiliation{Max Planck Society Outstation at the Heinz Maier-Leibnitz Zentrum (MLZ), D-85748 Garching, Germany}
\author{G.~Christiani}
\affiliation{Max-Planck-Institut f\"ur Festk\"orperforschung, Heisenbergstra\ss e 1, D-70569 Stuttgart, Germany}
\author{G.~Logvenov}
\affiliation{Max-Planck-Institut f\"ur Festk\"orperforschung, Heisenbergstra\ss e 1, D-70569 Stuttgart, Germany}
\author{M.~Kupriyanov}
\affiliation{Skobeltsyn Institute of Nuclear Physics, 119991, Moscow State University, Moscow, Russia}
\affiliation{Solid State Physics Department, Kazan Federal University, 420008 Kazan, Russia}
\author{A.~Sidorenko}
\affiliation{Institute of Electronic Engineering and Nanotechnologies ASM, MD2028, Kishinev, Moldova}
\author{B.~Keimer}
\affiliation{Max-Planck-Institut f\"ur Festk\"orperforschung, Heisenbergstra\ss e 1, D-70569 Stuttgart, Germany}
\date{\today}

\begin{abstract}
We report an investigation of the structural and electronic properties of hybrid superconductor/ferromagnet (S/F) bilayers of composition Nb/Cu$_{60}$Ni$_{40}$ prepared by magnetron sputtering. X-ray and neutron reflectometry show that both the overall interfacial roughness and vertical correlations of the roughness of different interfaces are lower for heterostructures deposited on Al$_2$O$_3$(1$\bar{1}$02) substrates than for those deposited on Si(111). Mutual inductance experiments were then used to study the influence of the interfacial roughness on the superconducting transition temperature, $T_C$. These measurements revealed a $\sim$4\% higher $T_C$ in heterostructures deposited on Al$_2$O$_3$, compared to those on Si. We attribute this effect to a higher mean-free path of electrons in the S layer, caused by a suppression of diffusive scattering at the interfaces. However, the dependence of the $T_C$ on the thickness of the ferromagnetic layer is not significantly different in the two systems, indicating a weak influence of the interfacial roughness on the transparency for Cooper pairs.
\end{abstract}

%\pacs{72.10.Fk, 74.25.F-, 74.45.+c, 74.62.Bf, 74.78.Fk, 73.50.-h}

\maketitle

\section{Introduction}
The interplay of two antagonistic states - superconductivity (S) and ferromagnetism (F) - at  interfaces has been in the focus of experimental and theoretical investigation for the past two decades \cite{Izyumov.Uspekhi2002,Buzdin.RMP.77.935.2005,Efetov.SpringerTracts.227.2007}. Theoretical work describes and predicts numerous manifestations of the mutual influence between the superconducting and ferromagnetic order parameters. These effects appear due to the oscillating nature of the decaying superconducting correlations in ferromagnetic materials that are proximity-coupled with the superconductor. One of the most striking manifestations of the interaction between both order parameters is a non-monotonic dependence of the critical temperature for superconductivity, $T_C$, in S/F heterostructures on the thickness of the ferromagnetic layer, $d_F$  \cite{Buzdin.RMP.77.935.2005,Zdravkov.PRL.97.057004.2006,Zdravkov.PRB.82.054517.2010}. The key parameter controlling the shape of the $T_C$($d_F$) dependence is the S/F interface transparency \cite{Fominov.PRB.66.014507.2002,Faure.PhysRevB.73.064505}, $T_{SF}$. For highly transparent ($T_{SF}\rightarrow1$) S/F interfaces, the $T_C$($d_F$)  curve oscillates, whereas for low ($T_{SF}\rightarrow0$) or moderate ($T_{SF}\lesssim0.5$) transparency it decays monotonically. To date, experimental research on S/F hybrids has mainly focused on the dependence of parameters characterizing the superconducting state (including $T_C$ and the critical current) on the thickness of the ferromagnetic layer, whereas the effect of the interfacial structure has remained practically unexplored. However, the characteristics of the interface may well be crucial for a detailed comparison with theoretical predictions and for the performance of functional S/F devices.

In the present paper, we address this issue by studying how the roughness of standard substrates used in modern vacuum technology (silicon and sapphire) influences the electronic proximity effect in S/F bilayers. We have used neutron and X-ray reflectometry as effective characterization methods of the structural properties of the layers and interfaces. Whereas specular reflectometry is a widely used method to determine the thicknesses of F and S layers and the root-mean-square (rms) roughness of the S/F interface \cite{Muehge.PRL.77.1857.1996,Muehge.PhysC296,Muehge.PRB.57.5071.1998,Lazar.PRB.61.3711.2000,Garifullin.ApplMagnRes.22.439.2002,Sidorenko.AnnPhys.12.37.2003,
Cirillo.PhysRevB.72.144511}, off-specular (diffuse) X-ray and neutron scattering is only rarely applied to S/F hybrid structures \cite{Tesauro.SuppercSciTech.18.1.2005,Nikitin.CrystRep.56.858.2011,Vecchione.SufSci.605.1791.2011}.
Off-specular scattering reveals further important characteristics of the interfaces, including the in-plane and vertical correlation lengths of the roughness profiles. In Nb/Cu systems prepared by molecular beam epitaxy (MBE) and by sputtering, X-ray off-specular reflectometry showed that the statistical properties of the roughness strongly depend on the deposition technique \cite{Tesauro.SuppercSciTech.18.1.2005}. In particular, samples grown by sputtering are characterized by roughness with high vertical correlation. Vertical roughness correlations were also reported for magnetron-sputtered [Nb/PdNi]$_{5-9}$ multilayers grown on Si(100) substrates \cite{Vecchione.SufSci.605.1791.2011}. This work demonstrated growth of the rms roughness with increasing number of deposited layers.

In present work, we used X-ray (XRR) and neutron (NR) reflectometry to characterize the structural properties of CuNi/Nb bilayers fabricated on silicon and sapphire substrates in a single deposition run. The resulting information yields important insights into the electronic properties. In particular, the rms height and in-plane correlation lengths of the interfacial roughness define the probability of diffuse scattering not only for neutrons and photons, but also for the electrons responsible for the proximity effects.  We discuss the consequences of these effects for the superconducting $T_C$ and its dependence on $d_F$.

The paper is organized as follows. Section II gives a short introduction to reflectometry. The sample preparation and the experimental setups are described in Section III. Section IV discusses the structural and electronic properties, and Section V provides a conclusion.

\section{X-ray and neutron reflectometry}

\begin{figure}[htb]
\centering
\includegraphics[width=\columnwidth]{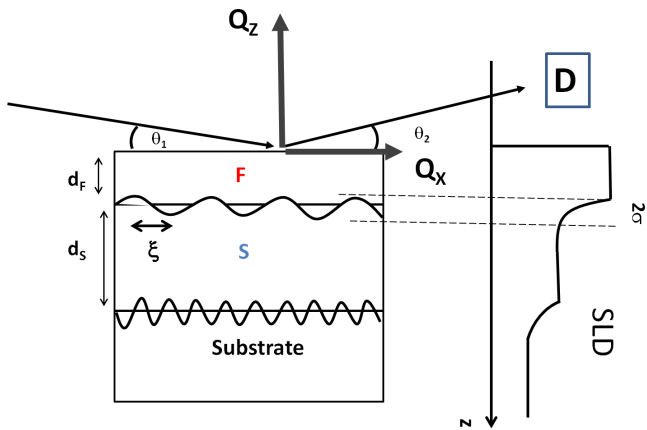}
\caption{(Color online) Scheme of the reflectometry experiment on a S/F
system. Rough interfaces are shown with wavy lines. Inplane
correlation length $\xi$ and rms height of roughness $\sigma$
are the typical length and height of the roughness profile at
certain interface. Corresponding scattering length density
(SLD) depth profile is shown on the right side of the
picture.}
\label{Fig1}
\end{figure}

Figure \ref{Fig1} illustrates the basic scheme of our reflectometric experiments. A beam of intensity $I_0$ with wavelength $\lambda$ falls on a sample under the grazing incidence angle $\theta_1$ and scatters under angle $\theta_2$. The detector \emph{D} is used to measure the intensity of the scattered beam $I_s(\theta_2)$. For flat layered structures with properties changing only along z-axis, the incoming beam is scattered only if the specular condition ($\theta_1 = \theta_2$) is satisfied.

In the kinematical approximation, the complex amplitude of specular reflection can be written as
\begin{equation}
r(Q_z) = \frac{4\pi}{Q_z^2}\int e^{i Q_z z} \frac{\partial}{\partial z} \rho(z) dz
\label{eq.1.r.Q}
\end{equation}
where $Q_z = 4\pi\sin(\theta_1) / \lambda$ is the momentum transfer in the direction normal to the sample surface, $\rho = N \times b_{\text{coh}}$ is the scattering length density (SLD), that is, the product of the atomic density $N$ and the coherent scattering length $b_{\text{coh}}$.
%Since neutrons scatter from nuclei, whereas X-rays scatter primarily from electron shells, the $b_{\text{coh}}$ for neutrons and X-rays differ significantly. %For the X-rays $b_{\text{coh}}$ is proportional to atomic number, while $b_{\text{coh}}$ for neutrons varies randomly from element to element and even from %isotope to isotope.

For a single S/F bilayer with sharp interfaces depicted in Fig. \ref{Fig1}, the reflection amplitude (\ref{eq.1.r.Q}) can be re-written as
\begin{eqnarray}
r(Q_z) = \frac{4\pi}{Q_z^2} [ -\rho_F &+& (\rho_F - \rho_S) e^{i Q_z d_F} +  \nonumber \\
&&(\rho_S - \rho_0) e^{i Q_z (d_F + d_S)}]
\label{eq.2.r.Q}
\end{eqnarray}
where $\rho_F$, $\rho_S$ and $\rho_0$ are the SLDs for the F and S layers and the substrate, correspondingly. The exponentials in Eq. \ref{eq.2.r.Q} lead to so-called \emph{Kiessig oscillations} in the experimentally measured reflectivity
$R(Q_z) \equiv I_0/I_s = \left| r(Q_z)\right|^2$. These oscillations are caused by interference of waves reflected from different interfaces. The period of oscillation, $\Delta Q$, is related to the layer thickness as $\Delta Q = 2\pi/d$. Hence this method can be used for the direct determination of the thicknesses of the S and F layers. However, it follows from Eq. \ref{eq.2.r.Q} that the sensitivity of the reflectivity to a given interface depends on its SLD contrast. Since the S and F layers are metals, often with similar atomic numbers, the X-ray contrast at the S/F interface, $(\rho_F - \rho_S)$, is usually small. As a result, XRR is mostly sensitive to the entire thickness of the S/F bilayer, $d_{SF}\equiv d_S+d_F$. For neutrons, the contrast at the S/F interface is usually higher and therefore, neutron reflectometry can be used for the precise determination of the parameters of the S/F bilayer.

Expression (\ref{eq.2.r.Q}) is valid for flat-layered systems with sharp interfaces. Roughness or interdiffusion leads to the broadening of the interface region over a range $2\sigma$. In this case, the corresponding amplitude of reflection from this interface must be corrected by the Debye-Waller factor, $\exp[-(Q_z \sigma)^2/2]$, (see Fig. \ref{Fig1}).  The parameter $\sigma$ is often called rms height of roughness (or simply rms roughness), although the specular reflectivity cannot distinguish between roughness and interdiffusion.

To this end, diffuse scattering has to be analyzed as well. The intensity of diffuse scattering in the Distorted Wave Born Approximation (DWBA) \cite{Sinha.PRB.38.2297.1988} can be written as

\begin{eqnarray}
I(\theta_1, \theta_2) \propto \sum_l \sum_m &&[\Psi(\theta_2, z_l) \Psi(\theta_1, z_l) \Delta\rho_l]^* \nonumber \\
&\times& C_{lm}\Psi(\theta_2, z_m) \Psi(\theta_1, z_m)\Delta\rho_m
\label{eq3.DWBA}
\end{eqnarray}
where $\Psi(\theta_{1(2)}, z_{l(m)})$ is the amplitude of the wave function of the incident ($\theta_1$) or scattered ($\theta_1$) radiation at the $l$-th ($m$-th) interface, calculated for the 1D in-plane averaged SLD depth profile (distorted waves). $\Delta\rho_{l(m)} \equiv \rho_{l-1(m-1)} - \rho_{l(m)}$ is the SLD contrast at the $l$-th ($m$-th) interface, $C_{lm}(Q_x)$ is the Fourier transformation of the correlation function $C_{lm}(x,y) = \left\langle \delta z_l(0,0) \delta z_m(x,y)\right\rangle$, of the height fluctuations $\delta z$ at the $l$-th and $m$-th interface. The auto-correlation function $C_{l,l}(x,y)$ describes the statistical properties of the roughness profile on the $l$-th interface. For exponential decay of the real-space correlation function, this can be written as \cite{Deak.PhysRevB.76.224420}
\begin{equation}
C_{l,l}(Q_x) = \frac{2\pi(\sigma_l\xi_l)^2}{(1+(Q_x\xi_l)^2)^{3/2}}
\label{eq4.Cll}
\end{equation}
where $\xi_l$ is the in-plane correlation length, $\sigma_l$ is the rms height of roughness at the $l$-th interface, and $Q_x=(2\pi/\lambda)(\cos(\theta_2) - \cos(\theta_1))$ is the parallel component of the momentum transfer (Fig. \ref{Fig1}). Vertical correlations of the roughness profiles at the $l$-th and $m$-th interfaces can be taken into account using the out of plane (vertical) correlation length $A$ \cite{Ming.PRB.47.16373.1993}:
\begin{equation}
C_{l,m} \rightarrow C_{l,l} \, e^{-\left| z_l - z_m\right|/A}
\label{eq5.Clm}
\end{equation}
The two limiting cases $A\rightarrow 0$ and $A \rightarrow \infty$ describe completely non-conformal and conformal roughness profiles, respectively. In the case of non-conformal roughness, the diffuse scattering from different interfaces adds incoherently:
\begin{equation}
I \propto \sum_l \left| \Psi(\theta_2,z_l)\right|^2 \left| \Psi(\theta_1,z_l)\right|^2 \left| \Delta\rho_l\right|^2
C_{ll}(\bm{Q_\parallel})
\label{eq6.I}
\end{equation}

As it follows from (\ref{eq4.Cll}) and (\ref{eq6.I}) for a single interface, the diffuse scattering intensity of quantum particles (neutron, photon or electron) is proportional to the product of the rms height and the correlation length of roughness, i.e. on the total volume affected by roughness. For multiple interfaces with conformal roughness, the intensity of the diffuse scattering comprises a coherent sum of contributions from every interface. Similar to Kiessig oscillations in specular reflection, the coherent sum of contributions from different interfaces generates an interference pattern in the diffuse scattering channel \cite{Holy.PRB.49.10668.1994}.

\section{EXPERIMENTAL DETAILS}
The samples were prepared in a magnetron vacuum machine (Leybold Z-400) on Si(111) and Al$_2$O$_3$(1$\bar{1}$02) substrates in a single deposition run at room temperature. The substrates were polished before the deposition to rms roughness of 0.2-0.3 nm and typical in-plane correlation length of several tens of microns. In total three targets were used: pure niobium (99.99\%) as a superconducting material, copper-nickel alloy (60\% Ni -40\% Cu) as a diluted ferromagnet, and pure Si (99.99\%) for growth of a protective capping layer. Sputtering was done under Argon (99.999\%, \emph{Messer Griesheim}) atmosphere of $8\times10^{-3} \text{ mbar}$ with a residual pressure in the chamber of about $1.5\times10^{-6} \text{ mbar}$. The design of the deposition machine allows growth of the entire structure in one cycle without disrupting the vacuum. This results in high-quality structures with smooth interfaces.
%\begin{table}[hb]
%\caption{\label{tab:table1} Nominal thickness of the prepared structures}
%\begin{ruledtabular}
%\begin{tabular}{ccccc}
%\textrm{No}&
%$d_S \text{[nm]}$&
%$d_F \text{[nm]}$&
%\textrm{substrate} &
%\textrm{buffer}\\
%\colrule
%1 & 8.0 & 15 & Si & $+$\\
%2 & 8.0 & 15 & Al$_2$O$_3$ & $+$\\
%3 & 20 & 10 & Si & $+$\\
%4 & 20 & 10 & Al$_2$O$_3$ & $+$\\
%5 & 10 & 0-40 & Si & $-$\\
%6 & 10 & 0-40 & Al$_2$O$_3$ & $-$\\
%\end{tabular}
%\end{ruledtabular}
%\end{table}
\begin{table}[hb]
\centering
\caption{\label{tab:table1} Nominal thickness of the prepared structures}
\begin{ruledtabular}
\begin{tabular}{ccccc}
\textrm{No} & $d_S \text{[nm]}$                  & $d_F \text{[nm]}$                    & \textrm{substrate} & \textrm{buffer}             \\
\colrule
1  & \multirow{2}{*}{8}  & \multirow{2}{*}{15}   & Si        & \multirow{2}{*}{$+$} \\
2  &                     &                       & Al$_2$O$_3$        &                    \\
3  & \multirow{2}{*}{20} & \multirow{2}{*}{10}   & Si        & \multirow{2}{*}{$+$} \\
4  &                     &                       & Al$_2$O$_3$        &                    \\
5  & \multirow{2}{*}{10} & \multirow{2}{*}{0-40} & Si        & \multirow{2}{*}{$-$} \\
6  &                     &                       & Al$_2$O$_3$        &
\end{tabular}
\end{ruledtabular}
\end{table}

In order to prevent contamination by absorbed gases and oxides, the targets and substrates were pre-sputtered for 3-5 minutes before deposition of the S/F structures. The deposition rates of the layers were $4.5\text{nm/s}$ for Nb, $3.5\text{nm/s}$ for CuNi alloy, and $1\text{nm/s}$ for Si. To obtain a homogeneous thickness of the deposited layers, the target was wobbled during the sputtering procedure. All samples were covered by silicon cap layers to protect them against oxidation.

Samples No. 1-4 (see Table 1) were prepared on $20\times20$ $\text{mm}^2$ substrates with buffer layers of silicon deposited in advance. Samples 5 and 6 were prepared using the wedge technology \cite{Zdravkov.PRL.97.057004.2006,Zdravkov.PRB.82.054517.2010} on $70 \times 7$ $\text{mm}^2$ substrates without silicon buffer layers. This technology allows the synthesis of samples with variable thickness of the F layer and constant thickness of the S layer, which can be used for precise determination of the $T_C$($d_F$) dependence. After deposition, these samples were cut into 31 (sample 5) and 28 (sample 6) pieces. The $n$-th piece of the $X$-th sample ($n$ = 5,6) will be referred to as $X$/$n$.

The NR- and XRR-experiments were conducted on the combined neutron/X-ray reflectometer NREX at the reactor FRM II (Technische Universit\"at M\"unchen). XRR profiles were measured using the Cu-K$_\alpha$  line with the wavelength $\lambda  = 1.54\AA$ . Neutron reflectivities were measured with a monochromatic beam with $\lambda =4.3\AA$ and divergence $0.02^{\circ}$. The superconducting transition temperature $T_C$  was determined by a custom-made mutual inductance setup by warming up the sample temperature from T $\approx$ 4 K with a rate $\sim 0.1\text{K/s}$. In this setup, samples were placed between the drive and pickup coils. The ac driving current was $60\text{mA}$ at a frequency $f$ = $1.6\text{kHz}$.

\section{EXPERIMENTAL RESULTS}

\begin{figure*}[htb]
\centering
\includegraphics[width=2\columnwidth]{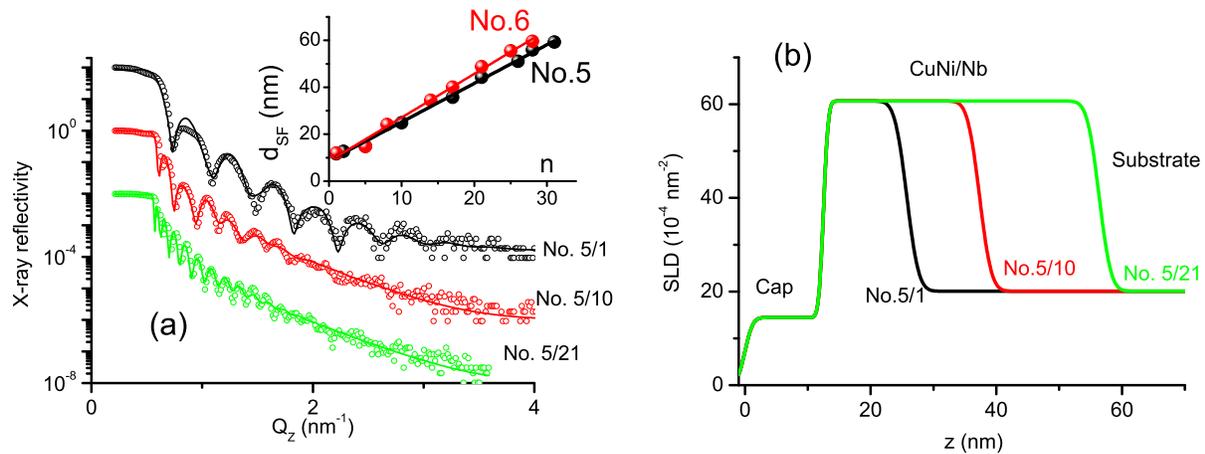}
\caption{(Color online) Experimental (dots) and model (solid lines) X-ray reflectivity curves for samples 5/1, 5/10 and 5/21. (b) SLD-depth profiles for samples 5/1, 5/10 and 5/21 corresponding to the solid lines in (a). Dots in inset to (a) shows dependence of fit-extracted thickness of S/F bilayer on piece number $n$. Solid lines are fit by linear dependence.}
\label{Fig2}
\end{figure*}

The specular X-ray reflectivities for samples 5/1, 5/10 and 5/21 are shown in Fig. \ref{Fig2}a. The curves are characterized by a single oscillation whose period grows with $n$. The presence of only one oscillation period in XRR is due to the negligible contrast (around 1\%) of the X-ray SLDs of the CuNi and Nb layers \cite{Khaydukov.JSupNovMag.28.1143.2015}. We fitted the experimental XRR data to model profiles, by varying the SLDs and thicknesses of the cap layer and the CuNi/Nb bilayer, as well as the rms roughness of the surface and the CuNi/substrate and Si/Nb interfaces. The SLD depth profiles resulting from these fits are shown in Fig. \ref{Fig2}b, and the dependence of the thickness of the S/F bilayer on $n$ is shown in the inset to Fig. \ref{Fig2}a. One can see that the thicknesses of the S/F bilayers for $n$=1 samples are similar ($d_{SF}=10.4$ nm), and that they grow linearly with coefficients 1.65 nm/$n$ and 1.86 nm/$n$ for samples 5 and 6, respectively.

\begin{figure*}[htb]
\centering
\includegraphics[width=2\columnwidth]{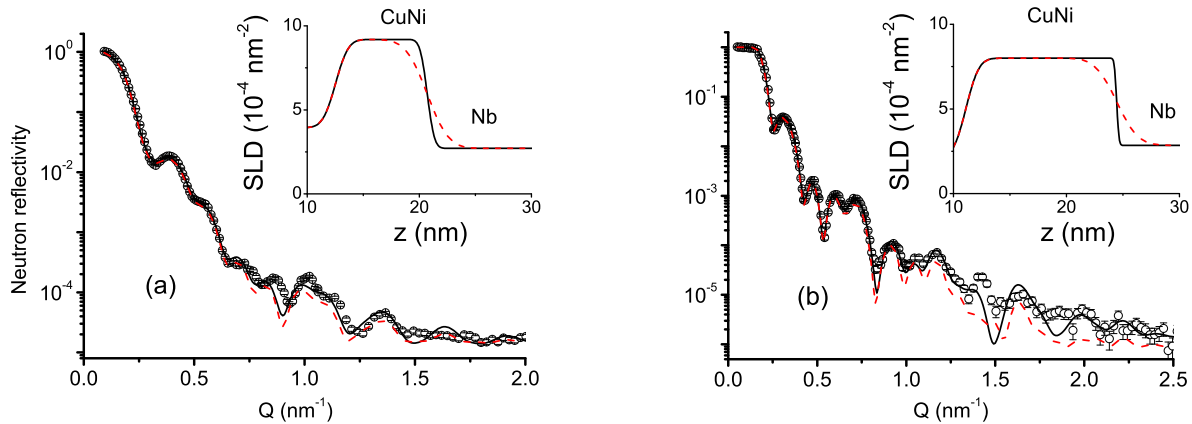}
\caption{(Color online) Experimental (dots) and model (solid lines) neutron reflectivity curves for samples 3 (a) and 4 (b). The inset shows the SLD depth profiles in the vicinity of the S/F interface. Dashed lines in insets and main graphs show SLD depth profiles with rough CuNi/Nb interface ($\sigma_{S/F} = 1.5 \text{nm}$) and corresponding reflectivity curves (all other parameters are the same).}
\label{Fig3}
\end{figure*}

Since the XRR is insensitive to the properties of the CuNi/Nb interface, additional neutron reflectometry measurements were performed on several samples. The neutron reflectivities measured on samples 3 and 4 are shown in Fig. \ref{Fig3}. One can see that sample 3 has more pronounced Kiessig oscillations due to the higher SLD of Al$_2$O$_3$ ($\rho_0=5.7\times10^{-4} \text{ nm}^{-2}$) compared to Si ($\rho_0=2.1\times10^{-4} \text{ nm}^{-2}$). Moreover, the reflectivity of sample 4 has an order of magnitude lower background level at high $Q_{Z}$ than the one of sample 3. The NR curves were fitted to model SLD depth profiles, according to the procedure described in Ref. \onlinecite{Khaydukov.JSupNovMag.28.1143.2015}. The best fit allowed us to extract the thicknesses of the S and F layers separately, as well as the rms height of roughness of the S/F interfaces, $\sigma_{S/F} = 0.5 \pm 0.1 \text{nm }$ (sample 3) and  $\sigma_{S/F} = 0.3 \pm 0.1 \text{nm }$ (sample 4). The results of the fit show that the rms height of the S/F interface roughness is higher for the sample 3 on the silicon substrate. To demonstrate the sensitivity of the NR curves to the rms roughness of the S/F interface, we present in Fig. \ref{Fig3} model curves for a rougher interface with $\sigma_{S/F} = 1.5 \text{nm}$. One can see that change of the $\sigma_{S/F}$ parameter leads to noticeable disagreement between model and experimental curves at $Q_z > 1\text{nm}^{-1}$.

The high neutron contrast between layers of CuNi ($\rho_F\approx8.5\times10^{-4} \text{ nm}^{-2}$) and Nb ($\rho_S\approx3.5\times10^{-4} \text{ nm}^{-2}$) allowed us to extract separately the thicknesses of the S and F layers. For the wedge sample 6/27, the fit of the NR data yielded $d_S = 11.3$ nm and $d_F = 44.6$ nm. Comparison with the XRR data shows that the total thickness of the S/F bilayer derived from NR and XRR agrees within 5\% accuracy. The thickness of the S layer extracted from the NR data also agrees approximately (within 8\%) with the value estimated from XRR data (see inset in Fig. \ref{Fig2}a).

\begin{figure*}[htb]
\centering
\includegraphics[width=2\columnwidth]{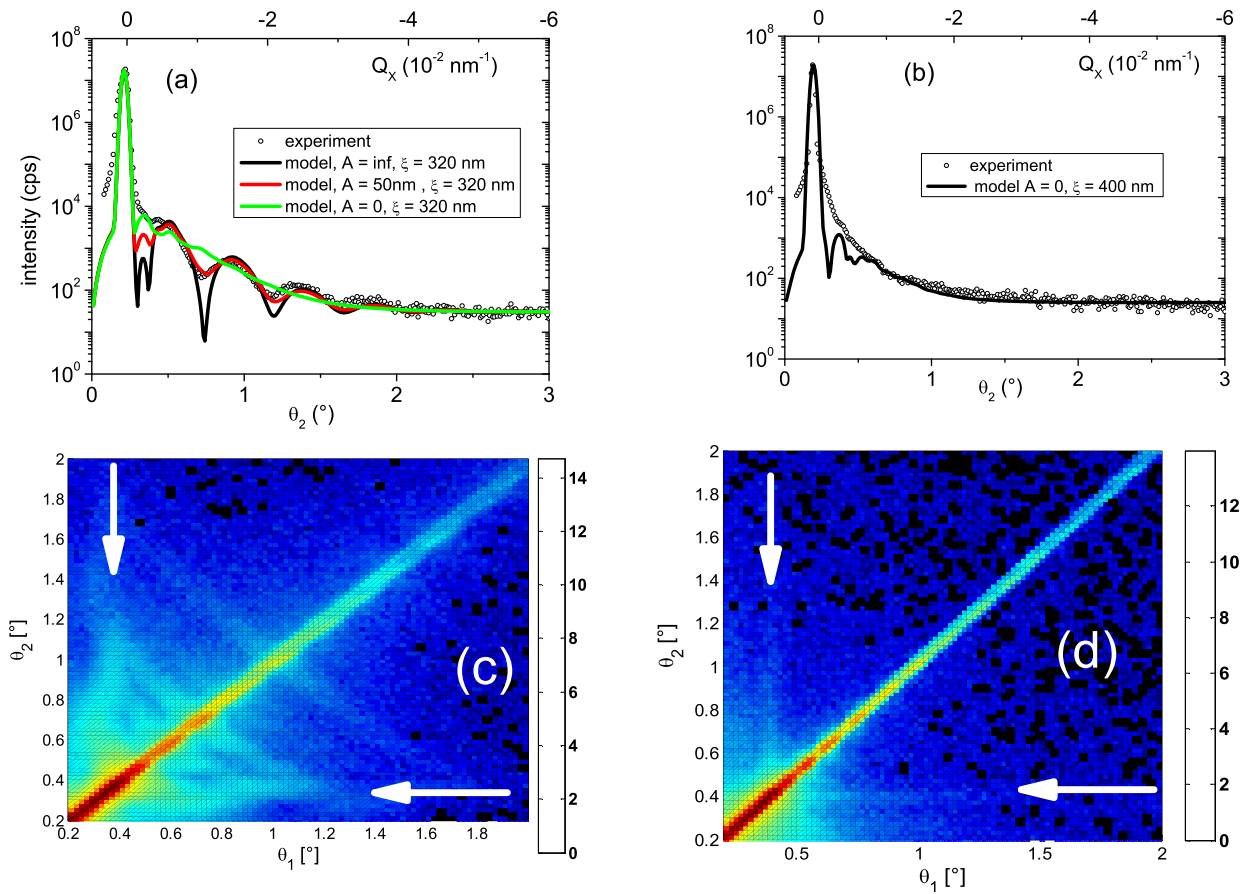}
\caption{(Color online) The X-ray detector scans measured on the samples 1 (a) and 2 (b) at $\theta_1 = 0.2^\circ$. Solid lines show the model curves for the different parameters. Experimental 2D scattering maps for the samples 3 (c) and 4 (d). Horizontal and vertical arrows show the positions of Yoneda scattering. Note that intensity in (c) and (d) is shown in ln-scale.}
\label{Fig4}
\end{figure*}

A more detailed study of the roughness profile was performed by analyzing the diffuse scattering intensity. Fig. \ref{Fig4}a,b shows X-ray detector scans for samples 1 and 2 measured at constant $\theta_1=0.2^\circ$, which exhibit a strong peak for specular reflection at $\theta_2 = \theta_1$ and a diffuse scattering tail 4-6 orders of magnitude lower in intensity. The diffuse scattering from sample 1 is characterized by pronounced oscillations (Fig. \ref{Fig4}a). As discussed in Section II, they originate from the interference of X-rays that are diffusely scattered from different interfaces. Changing the substrate to sapphire leads to the disappearance of these oscillations (Fig. \ref{Fig4}b).

The experimental scans were compared to model calculations performed within the DWBA. For calculations of the distorted waves, we used the SLD depth profiles derived from the fits to the specular XRR data. According to the fit, the rms roughness of the air/Si interface was 2 nm (0.6 nm),  Si/CuNi was 1.8 nm (0.4 nm) and Nb/buffer was 0.9 nm (0.5 nm) for sample 1 (2). Again, one can see that the rms roughnesses of the samples deposited on Si substrates are higher than those deposited on sapphire substrates.
%
%According to that fit  samples' 1 (2) air/capping layer, capping layer/CuNi and %Nb/substrate(buffer layer) interfaces are characterized by rms roughness of 2nm (0.6nm), %1.8nm(0.4nm) and 0.9nm(0.5nm).
The detector scans were fitted varying only in-plane $\xi$ and out of plane $A$ correlation lengths. The detector scan for sample 1 is well described by roughness with $\xi \simeq 320\text{nm}$. The presence of the oscillations is explained by the conformity of the roughness profiles of different interfaces with vertical correlation length $A \simeq 50 \text{nm}$, which is comparable to the total thickness of sample 1. To demonstrate the influence of the parameter $A$, we show in Fig. \ref{Fig4}a two curves for the limiting cases $A \rightarrow \infty$ and $A = 0$. One can see that for $A \rightarrow \infty$ Kiessig-like oscillations have much deeper minima than in the experiment. In contrast, for $A = 0$ the model curve demonstrates no oscillations, which is indeed the case for sample 2 (Fig. \ref{Fig4}b).  In that case the experimental data can  be described by a roughness with an in-plane correlation length $\xi \simeq 400\text{nm}$ without any vertical correlation (\emph{i.e.} $A = 0$).

Samples 3 and 4 exhibit closely similar behavior. The two-dimensional (2D) scattering maps for them are shown in Fig. \ref{Fig4}c,d. Both maps are featured by horizontal and vertical sheets of enhanced diffuse scattering around critical edge (so-called Yoneda scattering). The Yoneda scattering appears when either $\theta_1$ or $\theta_2$ is close to the critical edge and is explained by enhanced scattering on surface roughness \cite{Sinha.PRB.38.2297.1988}. The Yoneda scattering is also seen in Fig. \ref{Fig4}b,c as single peak close to the specular condition. The 2D  scattering map of sample 3 (Fig. \ref{Fig4}c) also shows tilted sheets of enhanced scattering intensity, which demonstrate conformity of the roughness profiles of different interfaces \cite{Holy.PRB.49.10668.1994}. Sample 4 grown on sapphire does not show this feature (Fig. \ref{Fig4}d). Comparison of Figs. \ref{Fig4}c and d (which were measured under identical experimental conditions) demonstrates that the total scattering intensity of sample 3 (deposited on the silicon substrate) scatters more strongly than sample 4 (on the sapphire substrate). The 2D pictures allow us to compare the intensities of specular and diffuse scattering for both samples quantitatively. The ratio of the diffuse and specular scattering intensities integrated over the entire measured region is 0.24\% and 0.15\% for samples 3 and 4, respectively. Samples on silicon substrates thus scatter more strongly in the off-specular direction than those deposited on sapphire.

The 2D scattering maps were also measured for neutrons. However, calculations based on the roughness parameters determined by off-specular XRR and SLD depth profiles from specular NR show that intensity of neutron diffuse scattering does not exceed the level of $10^{-6} I_0$, which is below the background level of the NREX reflectometer.

\begin{figure*}[htb]
\centering
\includegraphics[width=2\columnwidth]{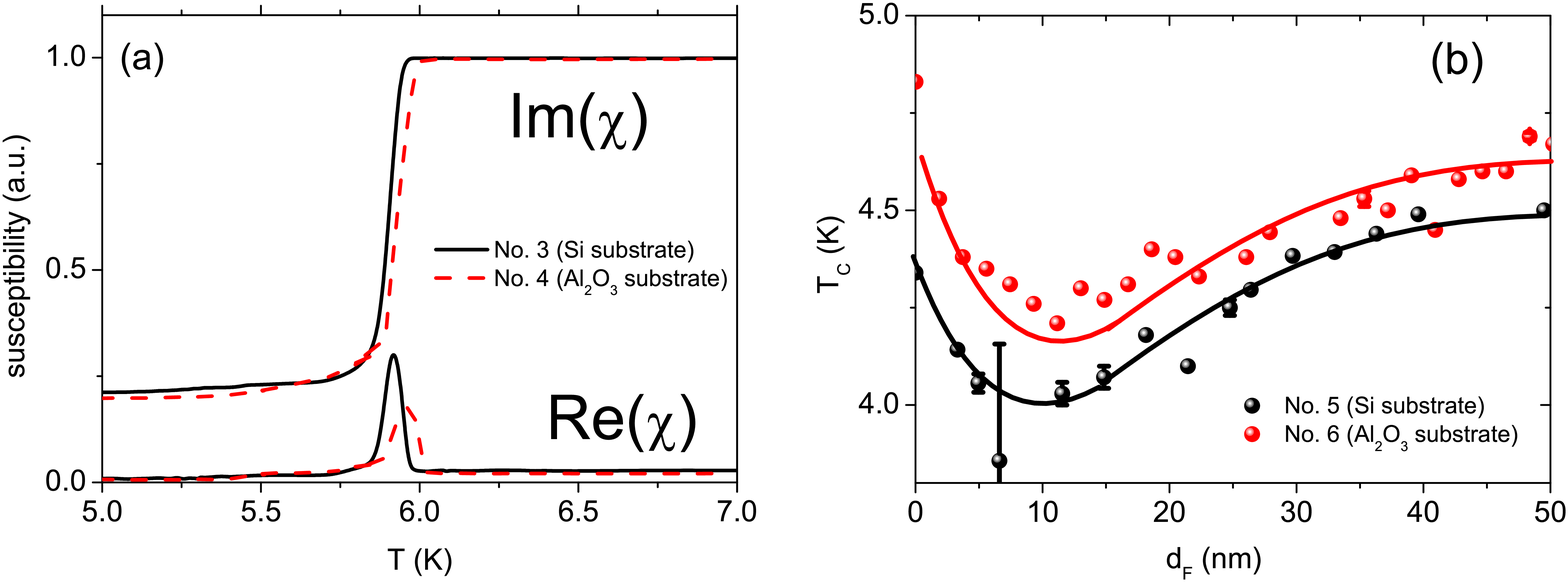}
\caption{(Color online) (a) Temperature dependence of the amplitude of the mutual inductance of samples 3 and 4. (b) Dependence of the superconducting transition temperature $T_C$ on $d_F$ for samples 5 and 6. The black solid line is a guide-to-the-eye. The red solid line is derived from the black one by multiplying with 1.04.}
\label{Fig5}
\end{figure*}

To study the possible influence of the interfacial roughness on the electronic properties and proximity effects, we have determined the superconducting transition temperature, $T_C$, of our samples in a mutual inductance setup (Fig. \ref{Fig5}). The temperature dependence of the real and imaginary parts of the measured susceptibility of samples 3 and 4 is shown in Fig. \ref{Fig5}a. Transition temperatures were defined as the midpoint of the transition of the imaginary part, which coincides with the center of the peak of the real part of the susceptibility. The transition temperature for the sample made on the silicon substrate $T_C = 5.92 \text{K}$ is in agreement with the data published previously \cite{Zdravkov.PRL.97.057004.2006, Zdravkov.PRB.82.054517.2010}. Sample 4 made on the sapphire substrate exhibits $T_C$ = 5.96 K which is $\sim$1\% higher than the value for sample 3. A more systematic study was performed on the wedge samples 5 and 6, which have the same $d_S$ and different $d_F$ (Fig. \ref{Fig5}b). The thicknesses of the S and F layers were derived from the comprehensive analysis of NR and XRR data, as described above. The $T_C$($d_F$) curve made on the Si substrate is similar to the previously reported data \cite{Zdravkov.PRB.82.054517.2010} which were described by calculations for a highly transparent ($T_{SF}\approx 0.6$) S/F interface. As follows from Fig. \ref{Fig5}b, the change of the substrate led to the systematic increase of $T_C$ for the all measured $d_F$. However, qualitative changes of the $T_C(d_F)$ curve were not observed. Curve for the sample 5 can be obtained from the curve for the sample 6 by scaling with a factor 1.04.

\section{DISCUSSION AND CONCLUSIONS}

The combination of specular and off-specular X-ray and neutron reflectometry has provided a comprehensive picture of the interfacial structure of CuNi/Nb bilayers grown on two different substrates. We first discuss the origin of the difference between the two systems. Our data showed that both the overall interfacial roughness and vertical correlations of the roughness of different interfaces are lower for heterostructures deposited on Al$_2$O$_3$(1$\bar{1}$02) substrates than for those deposited on Si(111). Presence/absence of the buffer layer does not influence on the conformity of the layers growth. Conformal growth has previously been observed for Cu/Nb/Cu trilayers \cite{Tesauro.SuppercSciTech.18.1.2005} and [Nb/PdNi]$_{5-9}$ periodic structures \cite{Vecchione.SufSci.605.1791.2011} prepared by sputtering techniques on Si(100) substrates. On the other hand, Cu/Nb/Cu/Si(100) and Cu/V/Fe/MgO(001) samples \cite{Nikitin.CrystRep.56.858.2011} grown by MBE did not show vertical correlation of roughness. To understand the reason of this behavior we must recall that the deposition rate of magnetron sputtering is one order of magnitude higher than that of the MBE process. Thus atoms sprayed onto a substrate by magnetron sputtering have a short time to distribute homogeneously over the surface. The spreading process is facilitated by the presence of vacancies on the substrate surface. If such spreading is not possible, the sputtered atoms repeat the roughness profile of a substrate, i.e. conformal growth of roughness takes place. The analysis of Si(111) and Al$_2$O$_3$(1$\bar{1}$02) surface made with the program VESTA \cite{jp.minerals.org} shows that the packing density of atoms on the Si(111) surface is two times higher than on Al$_2$O$_3$(1$\bar{1}$02). As a result, atoms being deposited onto the sapphire substrate have enough time to spread on the substrate, and correlations between the surface of the deposited layer and the in-plane profile of the substrate decay rapidly. By contrast, atoms deposited on the Si(111) surface spread more slowly, and the surface of the deposited layers thus repeats the substrate profile. As a consequence, the integrated strength of the diffuse X-ray scattering signal, which is a measure of the total volume affected by roughness, is also larger in the films grown on silicon.

The combined investigation of the structural and electronic properties of a large series of samples also allowed us to elucidate the origin of the variation of the superconducting transition temperature. Specifically, our data show that the interfacial roughness induced by deposition on different substrates influences only on the magnitude of $T_C$ and does not affect the oscillating character of the $T_C(d_F)$ dependence (Fig. \ref{Fig5}b). This implies that the transparency of the S/F interfaces is not noticeably affected by the interfacial roughness.  The 4\% difference in $T_C$ therefore cannot be attributed to differences in the proximity coupling to the adjacent F layer. In principle, the formation of an ultrathin metallic layer (silicides or oxides) between the Si substrate and the Nb layer \cite{Nakanishi.JApplPhys.77.948.1994,Zhang.JApplPhys.80.1422.1996} may also lead to a reduction of $T_C$ due to the proximity between the superconductor and normal nonmagnetic metal. We estimate that the presence of a metallic layer with a thickness of 0.1-0.2 nm could explain a 4\% suppression of $T_C$, as observed in our experiment (see Fig. 20 in Ref. \onlinecite{Golubov.JLowTPhys.70.83.1988}). However, a  related (albeit somewhat smaller) suppression of $T_C$ was also observed in samples 3 and 4, where the niobium layer was not in direct contact with the substrate, but instead with a silicon buffer layer. We also may exclude influence of the different lattice constants for Si and Al$_2$O$_3$ substrates, since previous studies have shown that the non-epitaxial growth of a niobium film goes in the direction (110) with spacing 0.23 nm not strongly depending on substrate (Si \cite{Tesauro.SuppercSciTech.18.1.2005}, Al$_2$O$_3$ \cite{Muehge.PhysC296} or glass \cite{Sidorenko.AnnPhys.12.37.2003}) or even deposition technique \cite{Tesauro.SuppercSciTech.18.1.2005}.

We therefore conclude that the difference of $T_C$ in the two systems primarily originates in the effect of interfacial roughness on the superconducting properties of the Nb layer, independent of its proximity to adjacent layers. Our data show that both the overall roughness and its vertical correlations are larger in the samples grown on Si. Diffusive electron scattering from rough interfaces is expected to reduce the electron mean free path in Nb layers deposited on Si relative to those deposited on sapphire. Previous work on superconducting films of various transition metals has shown that such diffuse scattering of electrons can reduce $T_C$, if the thickness of the layer is comparable to the superconducting coherence length, $\xi_S$  \cite{Strongin.Physica.55.155.1971,Teplov.JETP.1976}. The reduced mean free path smoothes out characteristic maxima in the electronic density of states, and hence reduces the density of states at the Fermi level $N(E_F)$. Indeed, for samples 5 and 6 with $d_S \simeq \xi_S \approx $ 10 nm \cite{Zdravkov.PRL.97.057004.2006,Zdravkov.PRB.82.054517.2010} the effect of  $T_C$-suppression reaches 4\%. For samples 3 and 4 with two times larger $d_S$ the suppression of $T_C$ is only 1\%.

In conclusion, the comprehensive characterization of the interfacial structure through complementary X-ray and neutron reflectometry experiments has allowed us to gain insight into the origin of differences in the electronic properties of S/F heterostructures grown on different substrates. The methodology we have described here may well prove more generally useful in research on electronic devices, where the influence of interfacial roughness and composition on the device performance is often poorly understood.

\begin{acknowledgments}

The authors are grateful to A. Senyshin (FRM II TUM) for assistance in the analysis of crystallographic data, and to C. Dietl (MPI FKF) and A. Kiiamov (Kazan Federal University) for assistance in mutual inductance and NR/XRR experiments. This work was carried out in the framework of a collaboration agreement between Max-Planck-Institut f\"ur Festk\"orperforschung and Institute of Electronic Engineering and Nanotechnologies ASM, and based upon experiments performed at the NREX instrument operated by Max-Planck Society at the Heinz Maier-Leibnitz Zentrum (MLZ), Garching, Germany. The neutron part of the project was supported by the European Commission under the 7th Framework Programme through the ``Research Infrastructures'' action of the Capacities Programme, NMI3-II, Grant Agreement number 283883. A. S. gratefully acknowledges financial support provided by JCNS to perform part of the neutron scattering measurements at the MLZ. R.M. and M.K. would like to thank support of Program of Competitive Growth of Kazan Federal University. This work was partially supported by the DFG collaborative research center TRR 80, IEEN Institutional research project No. 5982 and RFBR (Project 14-22-01007).
\end{acknowledgments}

\bibliography{CuNiBibliography}

\end{document}